# The effect of pulsed electromagnetic field exposure on osteoinduction of human mesenchymal stem cells cultured on nano-TiO$_2$ surfaces

Nora Bloise[1,2]*, Loredana Petecchia[3], Gabriele Ceccarelli[4], Lorenzo Fassina[5], Cesare Usai[3], Federico Bertoglio[1,2], Martina Balli[4], Massimo Vassalli[3], Maria Gabriella Cusella De Angelis[4], Paola Gavazzo[3], Marcello Imbriani[2,4], Livia Visai[1,2]*

1 Department of Molecular Medicine (DMM), Centre for Health Technologies (C.H.T.), INSTM Unit, University of Pavia, Pavia, Italy, 2 Department of Occupational Medicine, Toxicology and Environmental Risks, Istituti Clinici Scientifici Maugeri, IRCCS, Pavia, Italy, 3 Institute of Biophysics, National Research Council, Genova, Italy, 4 Department of Public Health, Experimental Medicine and Forensic, Centre for Health Technologies (C.H.T.), Human Anatomy Unit, University of Pavia, Pavia, Italy, 5 Department of Electrical, Computer and Biomedical Engineering, Centre for Health Technologies (C.H.T.), University of Pavia, Pavia, Italy

* nora.bloise@unipv.it (NB); livia.visai@unipv.it (LV)

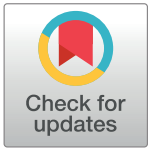

**Data Availability Statement:** All relevant data are within the paper and its Supporting Information files.

**Funding:** This study was supported by the Compagnia di San Paolo to CU, an INAIL grant entitled "Effetti dei campi elettromagnetici sulla salute umana: modelli sperimentali in vitro" (2011), COST Action grant BM1309 EMF-MED "European network for innovative uses of EMFs in biomedical applications" (2014–2019), and a COST Action

## Abstract

Human bone marrow-derived mesenchymal stem cells (hBM-MSCs) are considered a great promise in the repair and regeneration of bone. Considerable efforts have been oriented towards uncovering the best strategy to promote stem cells osteogenic differentiation. In previous studies, hBM-MSCs exposed to physical stimuli such as pulsed electromagnetic fields (PEMFs) or directly seeded on nanostructured titanium surfaces (TiO$_2$) were shown to improve their differentiation to osteoblasts in osteogenic condition. In the present study, the effect of a daily PEMF-exposure on osteogenic differentiation of hBM-MSCs seeded onto nanostructured TiO$_2$ (with clusters under 100 nm of dimension) was investigated. TiO$_2$-seeded cells were exposed to PEMF (magnetic field intensity: 2 mT; intensity of induced electric field: 5 mV; frequency: 75 Hz) and examined in terms of cell physiology modifications and osteogenic differentiation. Results showed that PEMF exposure affected TiO$_2$-seeded cells osteogenesis by interfering with selective calcium-related osteogenic pathways, and greatly enhanced hBM-MSCs osteogenic features such as the expression of early/late osteogenic genes and protein production (e.g., ALP, COL-I, osteocalcin and osteopontin) and ALP activity. Finally, PEMF-treated cells resulted to secrete into conditioned media higher amounts of BMP-2, DCN and COL-I than untreated cell cultures. These findings confirm once more the osteoinductive potential of PEMF, suggesting that its combination with TiO$_2$ nanostructured surface might be a great option in bone tissue engineering applications.

grant MODENA TD 1204, "Modelling Nanomaterial Toxicity" (2012–2016).

**Competing interests:** The authors received funding from the Compagnia di San Paolo. This does not alter our adherence to all the PLOS ONE policies on sharing data and materials.

## Introduction

The research on human mesenchymal stem cells from bone marrow (hBM-MSCs) has been an active field of investigation since 1970. Many studies assessed hBM-MSCs stability in culturing conditions and provided evidence of their immunomodulatory and tissue reparatory properties, selecting them as suitable candidates for many therapeutic applications, including improved healing of large bone defects, cell therapy and tissue regeneration. This great interest has emerged because of the multipotent ability of hBM-MSCs to naturally differentiate in several cell lineages, such as chondrocytes, adipocytes and osteoblasts. Noticeably, hBM-MSCs are the most susceptible to osteogenic differentiation among several populations of adult stem cells [1,2]. Cultivation of hBM-MSCs for regenerative purposes is a promising technique, but it requires special and expensive facilities to provide *in vitro* expansion to obtain an adequate number of cells to be implanted in the injured tissue.

Besides chemical agents, also physical factors, such as surface topography or external forces, proved to contribute in overcoming the drawbacks associated with standard culture systems and to improve their potential during *in vitro* culture. It is generally accepted that the surface topography (roughness, shape, and texture) of a biomaterial has an important effect on cellular attachment, adherence, proliferation and migration, as well as on the differentiation and survival of different cell types [3–5].

With respect to the bone, the creation of biomaterial surfaces with micro and nanoscale characteristics surely improves implants biocompatibility and osteointegration [6,7]. Currently, titanium dioxide (TiO₂) represents one of the most common and effective material for bone regeneration. In fact, the surface of TiO₂ can be modified to create a nanostructured surface matching native bone extracellular matrix (ECM) morphology and enhancing osteoblast incorporation and early osteointegration [4,8]. It has been observed that TiO₂ increases the adhesion of bone precursors, speeding up the osteogenic pathway activation [9,10]. In this context, we have recently shown that the growth of hBM-MSCs on TiO₂ nanostructured surface is a good approach to promote cell differentiation towards osteoblast lineage [11,12].

In literature there are interesting evidences that proliferation and differentiation of various cultured stem cells can also be increased by PEMF [9,13] Recently [14], we have characterized hBM-MSCs osteogenic differentiation with a special focus on Ca-related features of cell metabolism. We found that at least two Ca-pathways involved in the process of osteogenesis - namely the expression of L-type voltage gated Ca channels (VGCC) and the modulation of the concentration of cytosolic free $Ca^{2+}$ - were positively conditioned from repetitive exposure to low-frequency PEMF and thus can be proposed as reliable hallmarks of the osteogenic developmental stage.

hBM-MSC differentiation towards the osteoblastic lineage may be alternately mediated and promoted by different stimuli such as BMP-2 [15] and BMP-9 [16] or by the roughness of the growing surface [17]or by application of external forces [9,18].

In the present study, a well-established PEMF stimulation was applied on human bone marrow mesenchymal stem cells cultured on nanostructured TiO₂ substrates to investigate the effect of surface nano-topography in combination with exposure to low-frequency PEMF on cells differentiation, with special focus on alterations of $Ca^{2+}$-related aspects of cell metabolism.

## Material and methods

### Cell cultures

hBM-MSCs were isolated and phenotypically analysed to assess their mesenchymal properties according to the International Society for Cellular Therapy as previously described [13,14,19].

The study protocols were approved by the Institutional Review Board of the Fondazione IRCCS Policlinico San Matteo and the University of Pavia (2011). Written informed consent was obtained from all the participants enrolled in this study. The cells used in all experiments were mainly at passage 3. As described in our previous studies [11,14], hBM-MSCs were cultured at 37°C in a humidified incubator with 5% $CO_2$ in maintenance medium, low-glucose DMEM (Dulbecco's modified Eagle's medium) supplemented with 10% FBS, 1% glutamine, 50 µg/ml penicillin-streptomycin (P-S) and amphotericin B (Lonza Group Ltd.) (proliferative medium, PM). To induce osteogenesis, hBM-MSCs were cultured in osteogenic medium (OM), α-MEM (Minimum Essential Medium) supplemented with 10% FBS, 50 µg/ml of P-S and the osteogenic mixture containing 100 nM dexamethasone, 5 mM β-glycerophosphate disodium and 50 mg/ml ascorbic acid (Sigma-Aldrich, S. Louis, MO, USA). Treatment lasted up to 28 days and the medium was changed every 3 days.

## TiO₂ substrate

TiO₂ slides used in our experiments were the same as indicated in ref[11]. The scaffolds have been produced by Tethis S.p.a. (Milan Italy) and they are characterized by a granularity and porosity of few nanometres, mimicking an extracellular matrix environment and have been found to have a good degree of biocompatibility[20]. 20-mm diameter transparent coverslips loaded with a layer of transparent TiO₂ were used for the experiments.

## Pulsed electromagnetic field (PEMF)

The electromagnetic apparatus and the stimulation protocol was the same as in ref[14]: cells were exposed to PEMF for 10 min per day at the same time, adopting the following parameters: magnetic field intensity: 2 ± 0.2 mT; intensity of induced electric field: 5 ± 1 mV; frequency: 75 ± 2 Hz; pulse duration: 1.3 ms.

## Electrophysiology

Patch-clamp was utilized to perform current recording from the whole membrane of hBM-MSCs, as previously described [14]. At the beginning of treatment, hBM-MSCs were seeded into 12-well plates and cultured in OM to sub-confluence. Before recording, cells were detached from the substrate with trypsin-EDTA and resuspended in Standard Solution containing (values in mM): 150 NaCl, 5.4 KCl, 2.0 $CaCl_2$, 1.0 $MgCl_2$, 10 HEPES, 10 glucose (pH adjusted to 7.4 with NaOH). The suspension was stored at room temperature and used within few hours. Cells were transferred to the recording chamber and allowed to adhere to the glass bottom for 15 min. Subsequently, cells were continuously superfused by gravity flow (10 ml/min). Voltage activated $Ca^{2+}$ currents were measured in 108 mM $BaCl_2$ and 10 mM HEPES (pH 7.4, extracellular solution). Recording pipettes were in borosilicate glass and had tip resistance 3.0–5.0 MΩ when filled with a solution containing (values in mM): 8 NaCl, 40 KCl, 100 Aspartic Acid, 100 KOH, 2 $CaCl_2$, 5 EGTA and 4 adenosintriphosphate (ATP), 10 Hepes, pH 7.3. GEPULSE software was used for current acquisition and ANA [http://users.ge.ibf.cnr.it/pusch/programs-mik.htm] for current analysis. Further analysis was performed using Sigma Plot (SPSS Science, Chicago IL, USA).

## Epifluorescence imaging

To evaluate L-type $Ca^{2+}$ channel expression, cells were fixed in 4% PFA at room temperature, permeabilized with 0.2% Triton X-100 and blocked with 20% normal goat serum (Vector, Labs Burlingame, CA, USA) for 1 h at room temperature. Then cells were incubated with rabbit anti-human L-type -type $\alpha_{1C}$ subunit (CaV1.2) (1:100, Santa Cruz Biotechnology), washed and incubated with Alexa-fluor-488 (green) goat anti-mouse (Molecular Probes). Fluorescence was acquired with a Nikon Di-U upright microscope equipped with a Nikon DS-Qi1 digital CCD camera.

## Intracellular calcium measurement

To measure intracellular calcium, hBM-MSCs grown in PM or OM on TiO₂ substrate with/ without PEMF stimulation (as above described) were loaded with Fura-2 AM in the presence of Pluronic F-127 (Sigma-Aldrich, St. Louis, MO, USA) for 45 min at 37˚C. Measurements and data analysis were performed as previously described [12,14].

## Cell viability assay

To evaluate viability of hBM-MSCs seeded on TiO₂ nanosurface, a test with 3-(4,5-dimethylthiazole-2-yl)-2,5-diphenyl tetrazolium bromide (MTT; Sigma-Aldrich) was performed at day 7, 14, and 28 on cells cultured in proliferative medium or in osteogenic medium with/without PEMF stimulation as described in ref [14].

## qRT-PCR

Total RNA from hBM-MSCs seeded on TiO₂ nanosurfaces and cultured in proliferative or osteogenic medium with/without PEMF stimulation for 7 and 28 days, respectively, was extracted and retro-transcribed into cDNA as previously reported [13]. Gene expression analyses were performed by RT-qPCR using oligonucleotide primers displayed in S1 Table. Gene expression was analysed in triplicate and normalized to glyceraldehyde 3-phosphate dehydrogenase (GAPDH) gene expression.

## Confocal laser scanning microscopy (CLSM)

Samples in all experimental conditions were fixed with 4% (w/v) PFA for 1 h at 4˚C, washed with PBS three times for 15 min and blocked with PAT (PBS containing 1% [w/v] BSA and 0.02 [v/v] Tween 20) for 2 h at room temperature [14,21]. Anti-type I collagen, anti-osteocalcin, anti-osteopontin, and anti-alkaline phosphatase (ALP) rabbit polyclonal antisera (provided by Dr. Larry W. Fisher, National Institutes of Health, Bethesda, MD, USA) were used as primary antibodies diluted 1:500 in PAT. The incubation with primary antibodies was prolonged overnight at 4˚C and the negative controls were incubated with PAT alone. After washing, samples were incubated with Alexa Fluor 488 goat anti-rabbit IgG (Invitrogen, Carlsbad, CA, USA) at a dilution 1:750 in PAT for 1 h at room temperature. Cells were counterstained for 5 min with a solution of Hoechst (2 µg/mL; Sigma-Aldrich) to target nuclei and then examined under a confocal laser scanning microscope (CLSM) model TCS SP2 (Leica Microsystems, Bensheim, Germany).

## Bone ECM proteins extraction and ELISA assay

On day 28 of culture, an ELISA assay was performed as previously described to evaluate the amount of extracellular matrix proteins produced by unstimulated and PEMF-stimulated samples both cultured in proliferative or osteogenic medium [22]. Briefly, the samples were washed extensively with sterile PBS to remove culture medium and then incubated for 24 h at 37˚C with 1 mL of sterile sample buffer [20 mM Tris-HCl, 4 M GuHCl, 10 mM EDTA, 0.066% (w/v) sodium dodecyl sulphate (SDS), pH 8.0]. At the end of the incubation period, the total protein concentration was detected with the BCA Protein Assay Kit (Pierce Biotechnology

Inc., Rockford, IL, USA). Calibration curves to measure COL-I, COL-III, DCN, OPN, OSC, OSN, FN, BMP-2 and ALP were performed as previously described [22]. The amount of extra-cellular matrix constituents from each sample was expressed as pg/(cells × scaffold).

## ALP activity

ALP activity of hBM-MSCs seeded on TiO₂ and cultured in proliferative or osteogenic medium with/without PEMF stimulation was estimated at day 28 by a colorimetric assay as previously reported [21,22].

## Extracellular calcium deposition

In order to evaluate the calcium extracellular deposition calcium–cresolphthalein complexone method and fluorescent calcein detection were performed on hBM-MSCs seeded on TiO₂ and cultured in proliferative or osteogenic medium with/without PEMF stimulation at day 28 as previously described [13,23–26].

## Dot-blot assay

Dot blot analysis was performed to assess the amount of released-BMP-2, BOSP, DCN, OSN and COL-I in the culture media. Briefly, the culture media were collected every 3 days from unstimulated and PEMF-stimulated groups (both in PM and OM) up to 7 or 28 days of culture. Prior to dot blot analysis, each culture medium was lyophilized, and the final volume was adjusted to 50 μL in H₂O. Subsequently, 2 μL of the protein-medium extracts were spotted on a nitrocellulose membrane. After 1 h incubation in BSA 2%, samples were incubated with primary antibody diluted 1:1000 (Sigma Aldrich) at 4°C. Then, after 3 washes in TBS/Tween buffer (50 mM Tris/HCl, pH 7.4 containing 0.15 M NaCl, 0.05% Tween 20), the membrane was covered with secondary antibodies conjugated with HRP dissolved in BSA/TBS-T (1:500) for 1 h at room temperature. At the end, the membrane was finally incubated with ECL solution (GE Healthcare, Maidstone, UK) and viewed with ImageQuant LAS4000 (GE Healthcare). The spots were analysed with ImageQuant TL software (GE Healthcare) and then normalized to the number of cells counted with MTT assay at the time the culture media were collected.

## Statistics and data analysis

Each experiment reported in the Results Section was done in triplicates and at least in three separated experiments. Results were expressed as the mean ± standard deviation. All statistical calculation was carried out using GraphPad Prism 5.0 (GraphPad Inc., San Diego, CA, USA). Statistical analysis was performed using Student's unpaired t-test and one-way variance analysis (ANOVA), followed by *post hoc* Tukey test for multiple comparisons (significance level of $p \leq 0.05$).

## Results

hBM-MSCs cultured in osteogenic culture medium provided a clear increase in the differentiation efficiency either when seeded on TiO₂ coated substrates [11] or subjected to PEMF treatment[14]. Here the synergistic effect of these two factors was addressed to hBM-MSCs seeded on nanostructured TiO₂ substrates, unstimulated or PEMF-stimulated, in terms of calcium metabolism and osteogenic differentiation.

## Electrophysiology: VGCC properties and expression

Calcium currents activated from membrane potential were recorded from hBM-MSC population growing on TiO$_2$ in osteogenic medium (OM) and exposed or not exposed to PEMF (Fig 1). Ionic currents were elicited by applying voltage pulses 100 ms long from −40 mV to +80 mV, starting from a holding potential of −90 mV. In Fig 1A, Ca current traces from hBM-MSCs grown on TiO$_2$ in OM and exposed to PEMF for 28 days were recorded in standard solution (Ba 108.8 mM) and then in the presence of 10 µM nifedipine. The relative current-voltage relationships depicted in Fig 1B show that nifedipine, a specific blocker of L-type Ca channels, reduces the peak value of the maximal current by about 40-50%; this effect is always detectable during osteogenesis and it is independent from PEMF exposure or growth substrate. Thus, as already reported for glass[14], the molecular identity of the channels underlying Ca current in hBM-MSC is not changing even when cells are growing on the nanostructured TiO$_2$ substrate in proliferative medium or during osteogenesis and exposition to PEMF. The first row of bar charts in Fig 1C and 1D represents the number of cells growing adherent to TiO$_2$, expressing a detectable level of dihydropyridine-sensitive Ca current (C) and the values of the maximal current (D); in both cases, values were normalized vs. glass-grown cells. Three different conditions are shown: MSCs growing in proliferative medium (PM) and exposed to PEMF (TiO$_2$/PM/PEMF+); MSCs growing in OM and exposed (TiO$_2$/OM/PEMF+) or not exposed (TiO$_2$/OM) to PEMF. The results further confirm the stimulating effect of TiO$_2$ on the expression of functional Ca channels (Fig 1C and 1D and ref[12]) with respect to poly-l-lysine treated glass. However, when cells growing on TiO$_2$ were simultaneously exposed to PEMF and treated with osteogenic medium (TiO$_2$/OM/PEMF+), a clear decrease in the number of cell expressing L-type channels and in the mean amplitude of Ca current was observed. The same data reported in Fig 1C and 1D were observed under a different point of view and presented in Fig 1E and 1F, with the purpose of highlighting solely the influence of PEMF exposure on VGCC assembly and functioning. Indeed, the number of VGCC expressing cells (Fig 1E) and the Ca current amplitude of hBM-MSCs under PEMF stimulation (Fig 1F) were both normalized to the unstimulated samples of hBM-MSCs in proliferative or in osteogenic condition over a period of 28 days. The results excluded that the synergy existing between TiO$_2$ and PEMF in hBM-MSC osteogenesis involved VGCCs. Moreover, even in late differentiation, only the amplitude of the Ca current seems to be stimulated by PEMF exposure (Fig 1F). The development of the expression at the protein level of L-type Ca channels was monitored by immunostaining at the end of osteogenesis, in the presence or absence of PEMF stimulation (Fig 1G). According to the current literature, L-type VGCC are normally expressed in all mature osteoblasts, but only in few percentages of hBM-MSCs (see Fig 2 of ref[14]). Indeed, as Fig 1G shows, a clear green fluorescence was evident in osteo-differentiated hBM-MSCs after 28 days of *in vitro* treatment and PEMF exposure (TiO$_2$/OM/PEMF+). No difference was detectable at this stage between TiO$_2$ adherent cells exposed or not exposed to PEMF stimulation (data not shown), confirming that nanostructured titanium is a suitable substrate to elicit proper osteomarker expression during differentiation.

## Intracellular calcium measurement

Cytosolic free calcium is known to be the main second messenger regulating cellular physiological processes. In this work we monitored the changes of the basal level of intracellular calcium concentration, [Ca$^{2+}$]$_i$, at the beginning (day 3) and at the end (day 28) of osteogenic differentiation of hBM-MSCs growing on nanostructured TiO$_2$ (Table 1). The concentration for undifferentiated hBM-MSC cells was estimated to be 63 ± 4 nM, in full accordance with previous results (ref[12,14]). Furthermore, [Ca$^{2+}$]$_i$ increased nearly to 300 nM after 28 days of

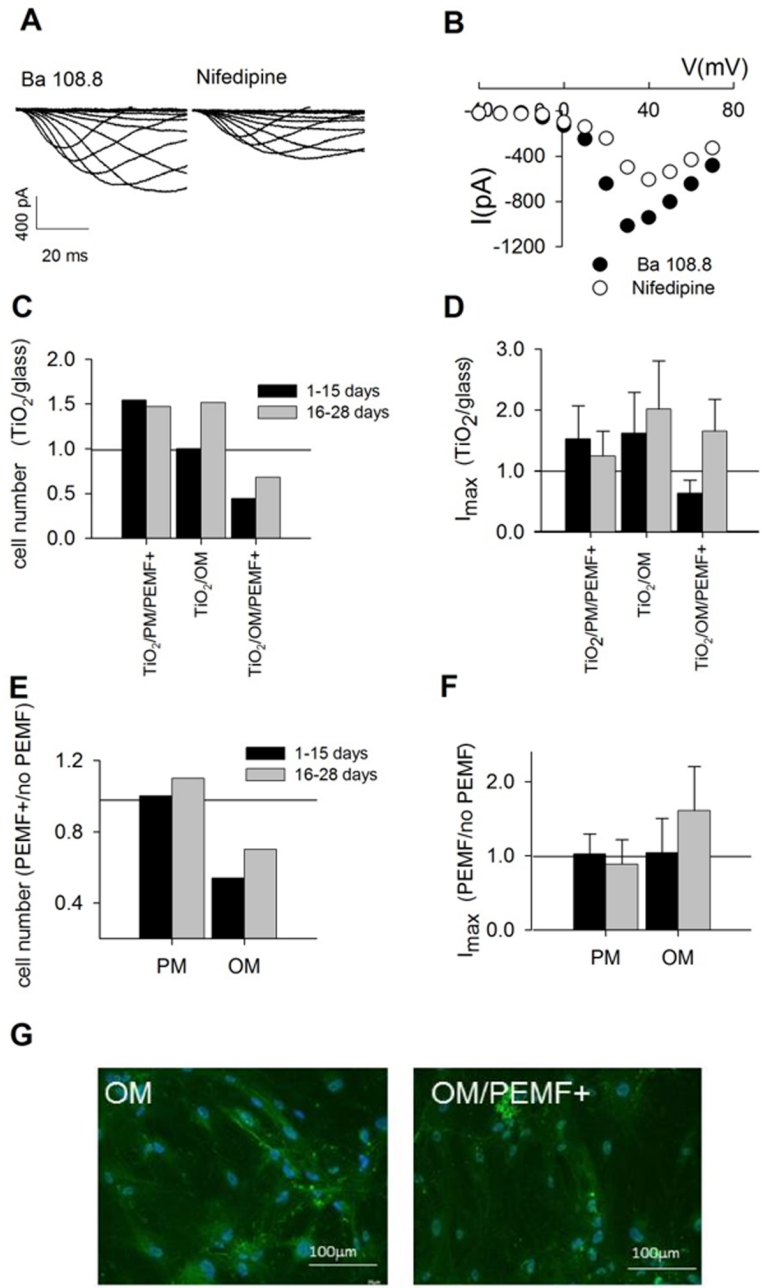

**Fig 1. PEMF exposure of hBM-MSCs grown on TiO₂ variously interferes in the expression of voltage-gated calcium channels (VGCCs).** A) Current traces were elicited applying voltage pulses 100 ms long from −40 to +70 mV. The amplitude of the current recorded in Ba 108.8 mM (left traces) decreased after the application of nifedipine, (right traces), a specific blocker of L-type VGCC. B) Current-voltage relationships of the traces in A. C) Number of MSCs growing on TiO₂ and expressing VGCCs during osteogenesis without/with PEMF exposure. Data were normalized to the number of cells grown on glass and subjected to the same treatment. D) Normalized VGCC $I_{max}$ value for the same population as in C. E) Number of cells grown on TiO₂ and exposed to PEMF normalized to the unstimulated sample and plotted in relation to the growing culture medium. F) Maximal current values of cell grown on TiO₂ and exposed to PEMF normalized to the unstimulated sample and plotted in relation to the growing culture medium. G) The expression of VGCC proteins carrying the currents as the one reported in panel A was confirmed by immunolocalization in MSCs grown on TiO₂ during osteogenesis (OM) and under PEMF stimulation (OM/PEMF+). No fluorescence was observed in hBM-MSCs grown in PM s a control (data not shown).

https://doi.org/10.1371/journal.pone.0199046.g001

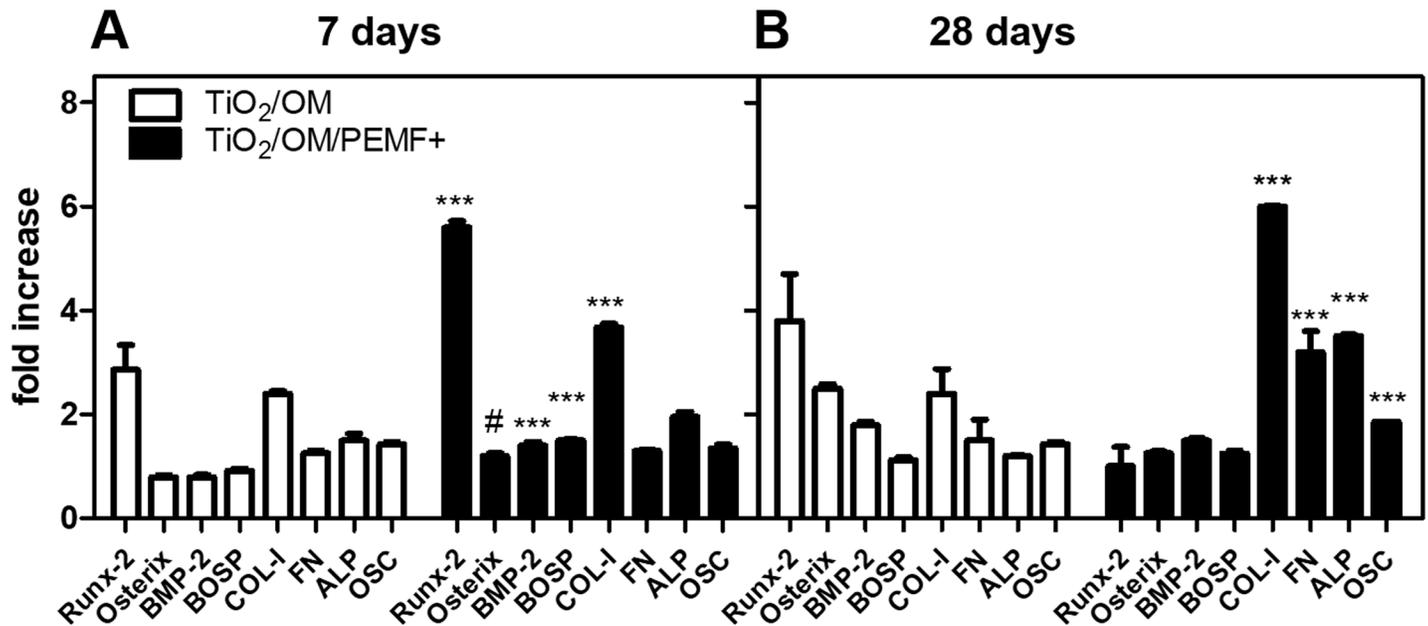

**Fig 2. Gene expression of the indicated bone-specific markers as determined by qRT-PCR.** hBM-MSCs were seeded and cultured in osteogenic medium on TiO₂ nanostructured surface with/without PEMF stimulation for 7 (A) and 28 (B) days, respectively. Statistical significance values are indicated as *** $p < 0.001$ and # $p > 0.05$.

osteogenic differentiation in cells growing on nanostructured titanium (TiO₂/OM)[12]. While the plateau values reached at the end of osteogenesis were independent from PEMF stimulation, the exposure to the electromagnetic field seems to speed up significantly the initial phase of the osteogenic process: $[Ca^{2+}]_i$ was 1.5-fold higher in PEMF-treated cells with respect to the untreated cells after only 3 days of OM treatment (155 ± 2 nM vs. 105 ± 2 nM, respectively). In absence of osteogenic compounds in the culture medium while maintaining the cells in proliferative condition, PEMF stimulated even more the rise of intracellular calcium in hBM-MSCs with an increase of 1.38-fold and 2.78-fold after 3 days and 28 days of culture, respectively (Table 1).

**Table 1. Intracellular concentration of Ca²⁺ ([Ca2+]ᵢ nM).**

| days | Proliferative medium (PM) | | Osteogenic medium (OM) | |
|---|---|---|---|---|
| | TiO₂ | TiO₂/PEMF+ | TiO₂ | TiO₂/PEMF+ |
| **3** | 63 ± 5 (§) (**) | 87 ± 2(# #) (***) | 105 ± 2(**) | 155±2 (***) (# #) |
| **28** | 69 ± 7 (§) (**) | 192 ± 8(# #) | 299 ± 11(**) | 293 ± 9(**) |

Intracellular concentration has been analysed performing Ca²⁺ Fura-2 measurements in hBM-MSCs. The trend of different cell populations growing adherent to TiO₂ was monitored after 3 days and at the end of the culture (28 days). The data, representing the results of three measurements in two separated experiments, are expressed as the mean ± SEM (** $p < 0.01$ vs. cultures grown on TiO₂ (§)). Similar values were also obtained in ref [12] even though from different experiments and at different checkpoints. Data represent the results of three sets of independent experiments and are expressed as mean ± SEM (** $p < 0.01$ and *** $p < 0.001$ vs. cultures grown on glass in PM; ## $p < 0.01$ vs. cultures grown on TiO₂).

## Osteogenic differentiation

In order to evaluate the synergistic action of PEMF and nanostructured titanium surface on osteogenic differentiation and extracellular matrix deposition, both untreated and PEMF-treated TiO$_2$ groups were examined by gene expression analysis, evaluation of ALP activity, calcium extracellular production, bone protein matrix deposition and secretion in culture media.

Firstly, the average cell viability was evaluated by MTT assay at different time frames (7, 14, 28 days) in cell population unstimulated and PEMF-stimulated daily either in proliferative or osteogenic culture media (S1 Fig). For all the marked checkpoints, the viability was in the range of 91%-98%, with no statistically significant difference between PEMF-stimulated and unstimulated samples (p > 0.05).

In agreement with our previous observations [11,14] in proliferative culture medium a slightly increase of selected osteogenic ECM protein deposition was detected for cells seeded on PEMF-treated TiO$_2$ surface in comparison with those unstimulated (S2 Table). For these reasons, we report samples of both untreated and PEMF-treated TiO$_2$ groups in the presence of osteogenic factors (OM).

qRT-PCR analyses were performed in cell samples collected at days 7 and 28 of culture, respectively, to define the cell phenotype (Fig 2). The oligonucleotides used as primers for qRT-PCR are reported in S1 Table, whereas, in Fig 2, the expression level of several genes is shown. The osteogenic genes expression in TiO$_2$/OM/PEMF+ groups significantly increased in comparison with TiO$_2$/OM groups after both 7 and 28 days of culture. At day 7, the expression of RUNX-2, COL-I and FN were significantly higher (3-fold, 2-fold and 2-fold, respectively) in comparison with unstimulated control (*** p < 0.001) (Fig 2A). At day 28 of culture, PEMF significantly increased the expression of BOSP (0.5-fold), Osterix (0.3-fold), OSC (0.5-fold), BMP-2 (0.2-fold), ALP (1.5-fold) in comparison to untreated control (*** p < 0.001) (Fig 2B).

Furthermore, the extracellular matrix constituents (ECM) were extracted and quantified by ELISA assay at the end of the culture to characterize the osteogenic process. As shown in Table 2, the detection of ECM proteins revealed some interesting differences. A significant enhancement in bone protein production was observed in PEMF-stimulated cells with an increment of 1.27- (ALP), 2.76- (COL-I), 1.44- (COL-III), 1.22- (FN) and 1.31- (OSC) fold in comparison with unstimulated cells (* p < 0.05). Conversely, a significant increase of OPN was determined in untreated cells than to PEMF-treated ones (* p < 0.05). Moreover, for DCN and OSN no significant differences were observed in the expression level between unstimulated and PEMF-stimulated cell cultures (# p > 0.05).

The production of COL-I, OSC and OPN was qualitatively visualized by confocal microscope analysis (Fig 3). Both in OM and OM/PEMF+ condition, cells showed a polygonal shape that resembling the osteoblast morphology may be consider a first evidence of the acquisition of osteogenic phenotype[27]. Moreover, the fluorescence analysis seemed to corroborate this observation. Upon exposure with PEMF, hBM-MSCs synthesized collagen-I, and started a process aims to externalize and assemble the type I collagen network as showed by a fluorescence distribution mainly at membrane level and in the interconnected spaces between the cells (Fig 3B). Without PEMF treatment, osteoblasts synthesized collagen, but the process of externalization seemed to be not yet completely activated or in an early stage as compared with PEMF treatment (Fig 3B). Regarding the non-collagenous proteins, immunofluorescence confirmed the production of both proteins but with small appreciable differences between untreated and treated samples. For instance, the presence in the peripheral region of cells of area positive to osteocalcin (Fig 3D) and as well the spotted-distribution of OPN (Fig 3F)

 

**Table 2. Matrix protein deposition after 28 days of cell culture in pg/(cell × scaffold).**

| Proteins | Osteogenic medium | | |
|---|---|---|---|
| | [a)]TiO₂ | [b)]TiO₂/PEMF+ | Ratio a/b |
| **ALP** | 18.94 ± 2.67 | 24.1 ± 2.80 | 1.27* |
| **COL-I** | 40.16 ± 4.00 | 111 ± 16.58 | 2.76* |
| **COL-III [a)]** | 66.28 ± 10.97 | 95.52 ± 0.12 | 1.44* |
| **DCN [b)]** | 86.13 ± 2.51 | 97.53 ± 7.94 | 1.13 |
| **FN** | 7.58 ± 2.30 | 9.28 ± 2.95 | 1.22* |
| **OPN [c)]** | 42.38 ± 5.08 | 23.59 ± 3.96 | 0.55* |
| **OSC** | 6.04 ± 0.29 | 7.96 ± 2.13 | 1.31* |
| **OSN [d)]** | 3.95 ± 0.21 | 3.39 ± 0.22 | 0.86 |

Protein quantification of bone ECM produced by hBM-MSCs cultured in osteogenic medium for 28 days on TiO₂ with/without PEMF stimulation. Results are expressed in pg/(cell × scaffold) and are presented as mean ± SD of three measurements in two separated experiments. In the table are also reported the ratio of TiO₂/OM vs. TiO₂/OM/PEMF+ (* p value < 0.05 vs. unstimulated samples).

[a)] COL-III, type-III collagen

[b)] DCN, decorin

[c)] OPN, osteopontin

[d)] OSN, osteonectin

In S1 Table were reported the abbreviations for ALP, COL-I, FN and OSC

https://doi.org/10.1371/journal.pone.0199046.t002

appeared slightly higher in OM/PEMF+ cells in comparison with the unstimulated cultures (Fig 3C and 3E).

Moreover, at the end of culture, ALP activity assessment showed a higher level of ALP activity in PEMF-stimulated samples in comparison to unstimulated control (* p < 0.05) (Fig 4A), in agreement with ALP immunostaining results (Fig 4B and 4C).

Furthermore, to determine the amount of calcium matrix deposition on unstimulated and PEMF-stimulated samples, calcium-cresolphthalein complexone protocol (Table 3) and fluorescent calcein staining (Fig 5) were also applied to samples from day 28. Interestingly, as reported in Table 3, we observed that TiO₂/OM group actively deposited a higher calcium content (almost 1.26-fold greater) in comparison to TiO₂/OM/PEMF+ cultures (* p < 0.05).

Similar trend was observed in the case of calcein staining, that binds the calcium mineral deposit in green. As clearly showed by CLSM images, a higher accumulation of green staining spots was observed in PEMF-treated cultures in comparison with those untreated (Fig 5).

Finally, to evaluate the content of BMP-2, BOSP, DCN, OSN and COL-I released in the culture media but not yet deposited from cells seeded on TiO₂ nanostructured surface, either unexposed or PEMF-exposed, a dot blot analysis was performed at days 7 and 28 (Fig 6). Significantly, on day 7 the levels of the specific bone proteins were too low to be detected by dot-blot assay, whereas interesting results were obtained at longer incubation time (Fig 6). At day 28, data showed that statistically significant differences were present in the culture media of PEMF-exposed cells with respect to the culture media of unexposed cells, for BMP-2 and DCN content (*** p < 0.01) followed by COL-I content (* p < 0.01). On the contrary, the levels of BOSP and OSN released in the culture medium were significantly lower in stimulated groups in comparison with the unstimulated control (** p < 0.001 and * p < 0.05 vs. untreated control for BOSP and OSN respectively).

From a general perspective, these results are in line with our previous studies (see ref [11,12,14]), thus they further point out osteogenic factors are essential elements for the development of a mature osteogenic phenotype for hBM-MSCs seeded on nanostructured TiO₂ surfaces and cultured in PEMF-treated condition.

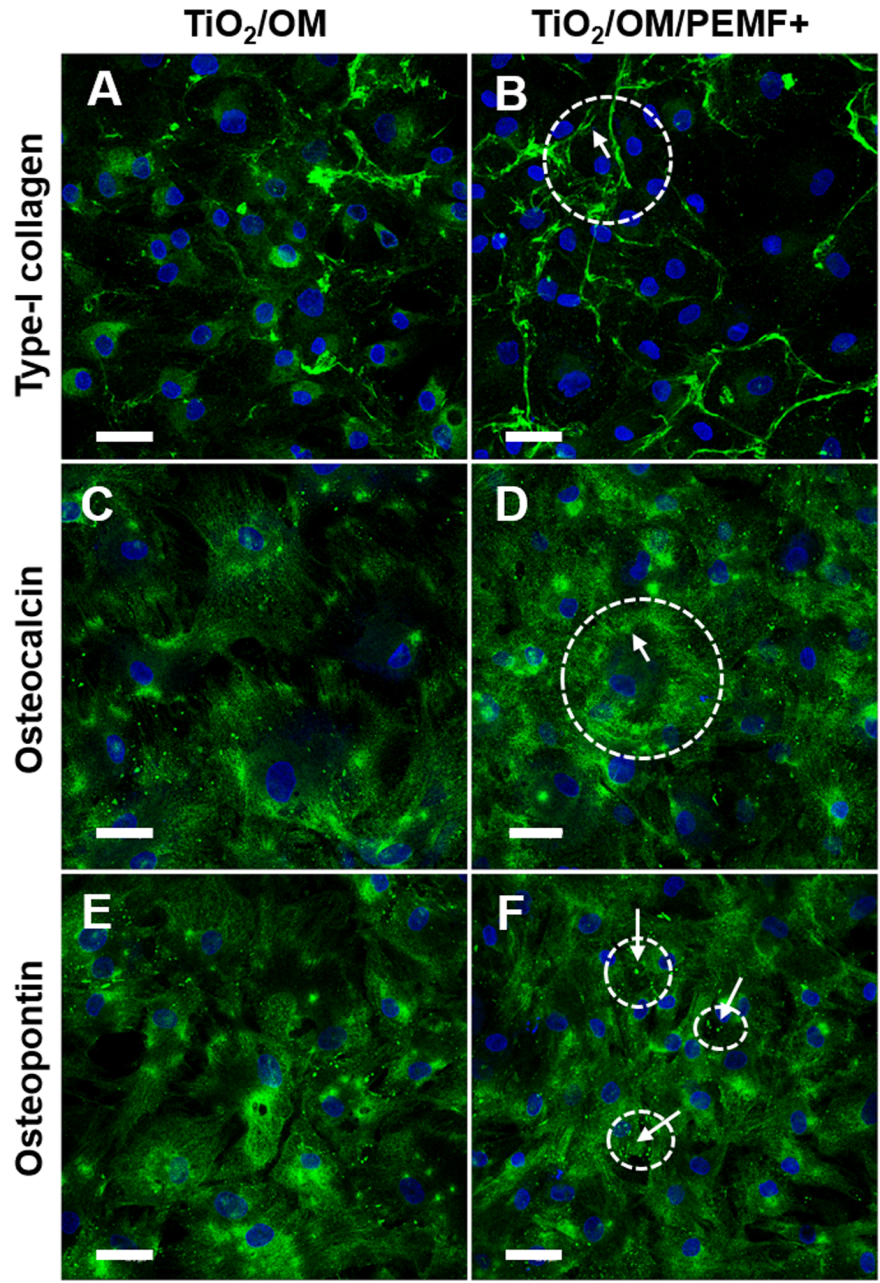

**Fig 3. CLSM images of bone proteins.** Immunolocalization of COL-I (A, B), OSC (C, D) and OPN (E, F) on cells cultured on TiO₂ surface in OM conditions for 28 days, unstimulated (A, C and E) or PEMF-stimulated (B, D and F). Arrows indicate the distribution of the immuno-stained proteins. Magnification 40X; the scale bar represents 50 μm. Nuclei (blue) were counterstained with Hoechst 33342.

## Discussion

For the regeneration of bone tissue, the most studied cell population is represented by MSCs (including bone marrow-derived mesenchymal stem cells (BM-MSCs), adipose tissue-derived stem cells (ADSCs), and dental pulp stem cells (DPSCs)) due to their ability to differentiate into bone forming cell types [28–31]. Despite the great potential of MSCs, there have been several studies that pointed out the importance of developing a method where biophysical and biochemical cues, through synergistic action, efficiently activate and support their osteogenic differentiation [32,33]. Indeed, increasing scientific evidence shows that the biophysical factors strongly affect stem cell properties. For example, the application of electrical field stimulation has been found to enhance the surface topography effects when determining orientation and elongation of fibroblasts and cardiomyocytes[34]. Additionally, *in vivo* experiments have demonstrated that low level laser therapy (LLLT) improves and accelerates bone repair within titanium scaffolds [35] and *in vitro* electromagnetic stimulation helps bone osteoblasts proliferation and extracellular matrix deposition on a titanium plasma-spray surface [36].

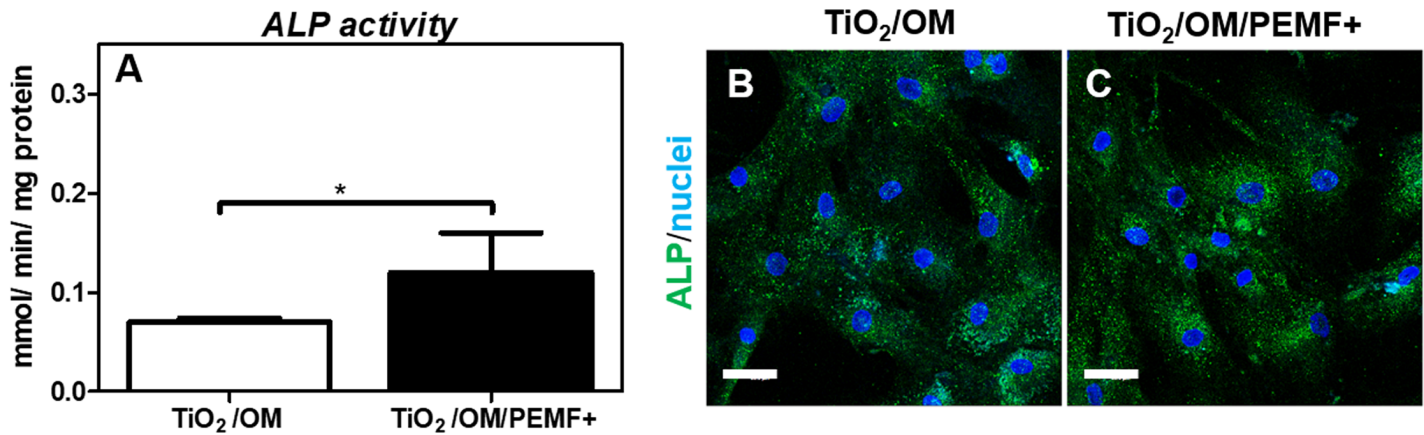

**Fig 4. ALP activity (A) and immunolocalization (B) of hBM-MSCs cultured for 28 days onto TiO₂ with/without PEMF stimulation.** A) ALP activity determined colorimetrically, corrected for the protein content measured with the BCA Protein Assay Kit and expressed as millimoles of $p$-nitrophenol produced per min per mg of protein. Bars express the mean values ± SEM of results from three measurements in two separated experiments (* $p < 0.05$). C). Immunolocalization of ALP following incubation with rabbit anti-human ALP primary antibody and detected with goat anti-rabbit secondary antibody (Alexa flour 488). Nuclei (in blue) were counterstained with Hoechst 33342. Magnification 40X; the scale bar represents 50 μm.

The aim of the present work was to address the biological response of human bone marrow derived mesenchymal stem cells cultured on nanostructured titanium dioxide surfaces in the presence of soluble osteoinductive factors and PEMF exposure. In comparison with unstimulated conditions, we demonstrated that PEMF-stimulated cells seeded on titanium nano-topography enhanced the biochemically induced osteogenesis of hBM-MSCs by the exploitation of molecular mechanisms that mainly interfere with some of the calcium-related osteogenic pathways, such as permeation and regulation of cytosolic and extracellular $Ca^{2+}$ concentration. Different papers support the idea that PEMF may be considered as a tool to improve autologous cell-based regeneration of bone defects in orthopaedics by stimulating the hBM-MSCs osteogenic differentiation [37,38].

Previously, we showed that PEMF stimulation - performed with the same protocol employed in this study - significantly increased the bone matrix deposition both in osteoblasts and in stem cells isolated from bone marrow and adipose tissue [13,36]. Moreover, we also

**Table 3. Extracellular calcium deposition in osteogenic medium expressed as ng $Ca^{2+}$/cell.**

| day | TiO₂ | TiO₂/PEMF+ |
|---|---|---|
| 28 | 4100 ± 0.1 | 5200 ± 0.05[(*)] |

Calcium deposition was analysed by calcium-cresolphthalein complexone method at day 28 of osteogenic culture condition. Results are expressed on a per-surfaces basis and presented as a mean ± SD of three measurements in two separated experiments (* $p < 0.05$ *vs.* cultures differentiated on TiO₂/OM).

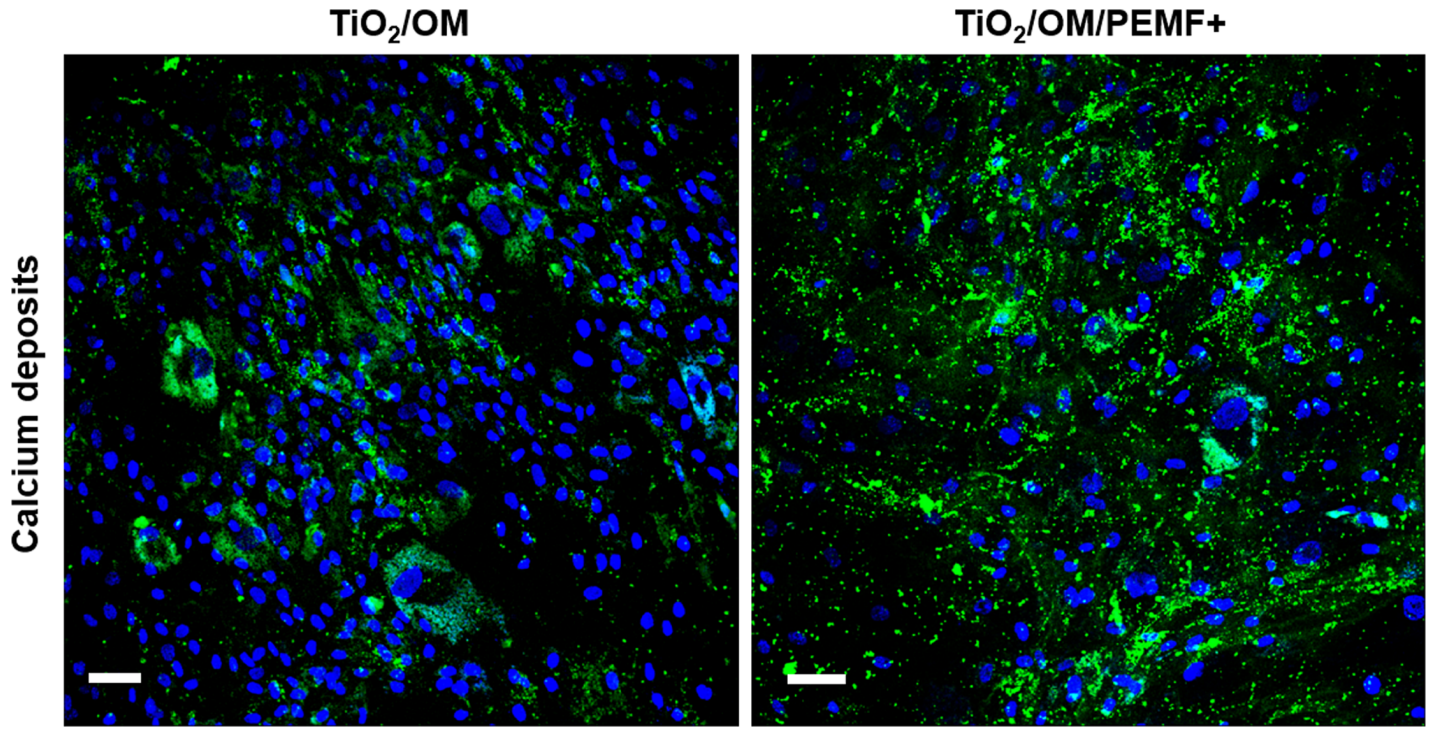

**Fig 5. Representative CLSM images of calcium deposits during the osteogenic differentiation after calcein staining.** Mineralized matrix regions were stained green with calcein and nuclei were stained blue with Hoechst 33342. Magnification 20X; the scale bar represents 50 μm.

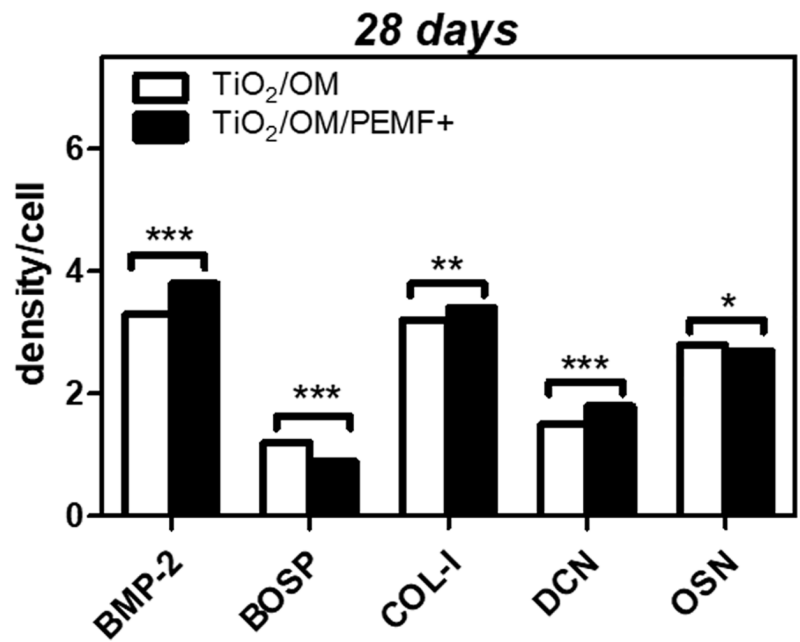

**Fig 6. Dot Blot of the indicated bone-specific proteins released in the osteogenic medium.** The culture medium of hBM-MSCs on TiO₂ with/without PEMF exposure was collected at day 28 and analysed as reported in the Material and Methods Section. Statistical significance values are indicated as $*$ $p < 0.05$, $**$ $p < 0.01$ and $***$ $p < 0.001$.

demonstrated the capability of $TiO_2$ nanostructured surface to promote hBM-MSCs differentiation to osteoblasts and its great potential in biomedical applications[11]. In this study, the combined effect of PEMF stimulation and nanostructured titanium on hBM-MSCs differentiation was evaluated. It has been demonstrated that PEMF affects osteoblast cell proliferation and differentiation to bone tissue by increasing DNA synthesis[39], but it also enhances, during cell differentiation, the expression of bone markers genes and the deposition of a calcified matrix [39]. Regarding osteogenic genes, we focused our attention on the expression at day 7 and at day 28 of both early (e.g. ALP) and late markers (e.g. osteocalcin) on hBM-MSCs in both culture conditions, unstimulated and PEMF-stimulated. We found that RUNX-2, the earliest marker of osteogenic commitment [40], showed a higher expression on $TiO_2$/OM/PEMF+ surfaces compared to the unstimulated one, endorsing the speeding up effects exerted by PEMF in osteogenesis. We detected higher expression levels of late-stage osteogenic differentiation genes (BOSP, OSC and ALP) on PEMF-stimulated cultures, another evidence in supporting the positive effect of the physical stimulus to increase the commitment of hBM-MSCs cultured onto $TiO_2$ towards osteoblasts, which resulted more effective in comparison to unstimulated $TiO_2$ cultures. The increase in osteogenic gene expression and calcium matrix deposition were consistent with the improvement in surface coating for PEMF-stimulated condition, showing the presence of proteins strictly involved in bone development, such as type I collagen, osteocalcin, osteopontin (as also confirmed by immunofluorescence observations). In particular, a higher amount of COL-I and OSC was determined in PEMF-stimulated cells, whereas OPN production resulted decreased in this condition. COL-I is the main constituent of the organic part of the ECM [41,42] whereas OPN and OSC are commonly used as early and late markers of osteogenic differentiation, respectively [30,41,43]. In fact, OPN is an extracellular glycosylated phosphoprotein secreted during the early stage of osteogenesis before the beginning of the bone matrix mineralization, whereas OSC secretion occurs after the onset of mineralization. Based on these findings, apparently, PEMF exposition by seems to accelerate cells toward the bone differentiation pathway influencing in a significant manner all the proteins involved in this process. Another relevant observation comes from qualitative immunofluorescence analysis. The distribution pattern of COL-I was significantly changed in PEMF-treated cultures than untreated ones. COL-I appeared mainly localized at the membrane level or in the extracellular matrix, that is a further evidence of PEMF capability as accelerating factor of the extracellular collagen production and bone mineralization. Concomitant with this collagen network formation in the PEMF-treated cultures, osteoblast-resembling cells showed a slight enhancement of small positive area in peripheral cellular regions for osteopontin and osteocalcin, both mineral-binding proteins, which presence was by its self indirectly indicative of mineralization process [41,43,44]. Indeed, it is well established that OPN and OSC closely associate with nascent and growing hydroxyapatite crystals in the extracellular matrix, thus regulating and supporting the formation of bone mineral tissue [44,45].

All these data, together with the rise of ALP expression and activity in PEMF-stimulated samples, seem to indicate that PEMF exposure drives a more rapid commitment of stem cells towards osteoblastic differentiation.

In order to corroborate and validate the results obtained by ELISA assay, a dot-blot analysis was performed on culture media from unstimulated and PEMF-stimulated cells at day 28. The possibility to measure some cytokines or osteogenic factors in culture media could be related to tissue-specific effects mediated via autocrine and paracrine actions[46]. It has been demonstrated that BMP-2 is an important cytokine that modulates and promotes osteogenesis by mediating the condensation of mesenchymal cells along an osteoblastic phenotype[47]. Our results demonstrated the secretion of extracellular matrix components in the medium of PEMF-treated cells, showing the highest amount of secreted BMP-2, DCN and COL-I; this phenomenon seems to indicate that PEMF stimulation significantly improved hBM-MSCs' capability to release bone proteins in the culture media. The presence of BMP-2 and COL-I in culture media could represent a possible paracrine mechanism elicited by pulsed electromagnetic field to promote transcriptional factors and other cytokines to direct the up-regulation of target genes and to fulfil their roles in differentiation of osteoblasts.

All the obtained results sustained the idea that the application of the electromagnetic exposure, associated to surface characteristics, provides an increase of the differentiation efficiency of hBM-MSCs towards the osteo-lineage. Nevertheless, the co-operative effect seems to influence the cell physiology in a manner, which is not a simple sum of the two separate actions. This is clearly exemplified by electrophysiology data on the expression of L-type VGCC channels. Looking at the quantification proposed in Fig 1, it is evident that, whereas TiO₂ nanostructure and PEMF stimulation are separately enhancing the expression of VGCC channels, the cumulative effect is providing the opposite outcome, even though the difference is reduced with time. This striking result is highlighting the fact that cells are interpreting the external stimuli, and a sort of biological feedback is activated, putatively reducing VGCC channels expression to prevent a potentially toxic, excessive rise of calcium concentration. The balance is thus going to be re-established in time and cells are going to recover a common phenotype when reaching the full differentiation.

In summary, based on electrophysiological, molecular and protein results, we may conclude that the co-application of chemical and physical factors is influencing the osteogenic differentiation of hBM-MSCs in a complex manner, which is not the simple sum of each isolated effect. Surface nanostructure, OM treatment and PEMF stimulation have been demonstrated to alter cellular calcium homoeostasis but the overall effect of an integrated treatment is strongly non-summative.

This result suggests the existence of an interplay of these external factors with yet unidentified intracellular calcium regulation mechanisms.

The main priority for future research should be to better clarify this point in order to design and establish reliable methods for stimulation and guidance of stem cells differentiation, to be exploited in regenerative and personalized medicine applications.

## Supporting information

**S1 Fig. Effect of pulsed electromagnetic field (PEMF) exposure on viability of human bone marrow derived mesenchymal stem cells (hBM-MSCs) seeded on TiO₂ surfaces.** Stem cells grown in proliferative (A) or osteogenic (B) medium and normalized to the unstimulated samples (TiO₂-seeded cells) cultured in the same culture medium (PM, A; OM, B). MTT cell viability assay was performed daily at different time frames of culture stimulating stem cells with PEMF. The control was represented by unstimulated culture (TiO₂-seeded cells). Data are presented as viability percentage to unstimulated culture set equal to 100%. Bars indicate mean values ± standard error of the mean of results from three experiments.
(PDF)

**S1 Table. Primers used for qRT-PCR study.**
(PDF)

**S2 Table. Matrix protein deposition after 28 days of cell culture in proliferative medium [expressed as pg/(cells × scaffold)].**
(PDF)

## Acknowledgments

The authors wish to thank P. Vaghi (Centro Grandi Strumenti, University of Pavia, Pavia, Italy, http://cgs41.unipv.it/wordpress/) for technical assistance in the CLSM studies and M. Vercellino for cell culture support. We are grateful to M.A. Avanzini (IRCCS, Policlinico San Matteo, Pavia, Italy) for mesenchymal stem cells isolation and characterization. We also thank M. Bordoni (University of Pavia) for reviewing the English.

## Author Contributions

**Conceptualization:** Massimo Vassalli, Paola Gavazzo, Livia Visai.

**Formal analysis:** Nora Bloise, Gabriele Ceccarelli.

**Funding acquisition:** Massimo Vassalli, Paola Gavazzo, Livia Visai.

**Investigation:** Nora Bloise, Loredana Petecchia, Gabriele Ceccarelli, Federico Bertoglio.

**Methodology:** Lorenzo Fassina.

**Supervision:** Massimo Vassalli, Paola Gavazzo, Livia Visai.

**Validation:** Nora Bloise, Loredana Petecchia, Gabriele Ceccarelli, Maria Gabriella Cusella De Angelis.

**Visualization:** Nora Bloise, Loredana Petecchia.

**Writing – original draft:** Nora Bloise, Massimo Vassalli, Paola Gavazzo, Livia Visai.

**Writing – review & editing:** Nora Bloise, Loredana Petecchia, Gabriele Ceccarelli, Lorenzo Fassina, Cesare Usai, Federico Bertoglio, Martina Balli, Massimo Vassalli, Maria Gabriella Cusella De Angelis, Paola Gavazzo, Marcello Imbriani, Livia Visai.